\documentclass{iaus}
\usepackage{graphicx}

\title[Direct Imaging of Planet Transit Events] 
{Direct Imaging of Planet Transit Events}

\author[G.T. van Belle et al.]   
{Gerard T. van Belle$^1$, Kaspar von Braun$^2$, Tabetha Boyajian$^3$, \and Gail Schaefer$^4$}

\affiliation{
$^1$Lowell Observatory, Flagstaff, Arizona, USA
\\ email: {\tt gerard@lowell.edu}
\\[\affilskip]
$^2$California Institute of Technology, Pasadena, California, USA\\email: {\tt kaspar@ipac.caltech.edu}
\\[\affilskip]
$^3$Yale University, New Haven, Connecticut, USA\\email: {\tt tabetha@chara.gsu.edu}
\\[\affilskip]
$^4$Georgia State University, Atlanta, Georgia, USA\\email: {\tt schaefer@chara-array.org}
}

\pubyear{2008}
\volume{xxx}  
\pagerange{119--126}
\jname{Title of your IAU Symposium}
\editors{A.C. Editor, B.D. Editor \& C.E. Editor, eds.}
\begin{document}

\maketitle

\begin{abstract}
Exoplanet transit events are attractive targets for the ultrahigh-resolution capabilities afforded by optical interferometers.  The intersection of two developments in astronomy enable direct imaging of exoplanet transits: first, improvements in sensitivity and precision of interferometric instrumentation; and second, identification of ever-brighter host stars.  Efforts are underway for the first direct high-precision detection of closure phase signatures with the CHARA Array and Navy Precision Optical Interferometer.  When successful, these measurements will enable recovery of the transit position angle on the sky, along with characterization of other system parameters, such as stellar radius, planet radius, and other parameters of the transit event.  This technique can directly determine the planet's radius independent of any outside observations, and appears able to improve substantially upon other determinations of that radius; it will be possible to extract wavelength dependence of that radius determination, for connection to characterization of planetary atmospheric composition \& structure.  Additional directly observed parameters - also not dependent on transit photometry or spectroscopy -  include impact parameter, transit ingress time, and transit velocity.
\keywords{stars: planetary systems, instrumentation: high angular resolution, instrumentation: interferometers}
\end{abstract}

\firstsection 
\section{Introduction}

Recent discoveries of stars exhibiting the telltale signs of planet transits has begun to add a new layer of understanding to the rapidly developing field of comparative exoplanetology.  The detected transit events have served to define the specific nature of those planets, including parameters such as density, atmospheric composition, and aspects of system dynamics (\cite[Burrows et al. 2006]{2006ApJ...650.1140B}).

Advances in the state of the art in astronomical optical interferometry can be directed at these recent transit discoveries and also contribute to telling the exoplanet story.  Specifically, interferometric observations during a planet transit event can determine the inclination and orientation of the planetary orbit upon the sky, in addition to refining the angular diameter measurements of both the planet and the star. Just as the Rossiter-McLaughlin effect (\cite[Rossiter 1924, McLaughlin 1924]{1924ApJ....60...15R,1924ApJ....60...22M}) in radial velocity measurements can contribute to our knowledge of transiting planet system parameters (\cite[Winn et al. 2006]{2006ApJ...653L..69W}), transit event interferometry can further the physical description of these systems, including measurements of the transiting planet's orbital plane orientation upon the sky.  These observations also uniquely determine the other observables of the system - impact parameter, transit velocity, stellar radius, planet radius, transit ingress time - without the need for supporting observations such as transit photometry.  For example, the previous direct determination of HD189733b's diameter (\cite[Baines et al. 2007]{baines2007}) measured that parameter through a combination of interferometric measurements and transit photometry; this technique is independent of such outside measurements.  A detailed evaluation of interferometric transit observations can be found in \cite{2008PASP..120..617V}.


\section{Instrument Capabilities}\label{sec_instruments}

Two instruments are currently available for attempting observations of exoplanet transits: the Michigan Infrared Combiner (MIRC) on the Center for High Angular Resolution Astronomy (CHARA) Array and the Visible Imaging System for Interferometric Observations at NPOI (VISION) instrument on the Navy Precision Optical Interferometer (NPOI). MIRC, provided to the CHARA Array by the University of Michigan, has a capability to combine 6 telescopes at H \& K and provides 10 simultaneous closure phase measurements on sources, representing a major step forward in capability for the facility.  The 6-way visible-light Tennessee State University VISION instrument (\cite[Ghasempour et al. 2012]{2012AAS...21944613G}) was recently deployed on NPOI, which is operated by science partners Lowell Observatory, the United States Naval Observatory (USNO), and the Naval Research Laboratory (NRL).  VISION saw first light in September 2012, and its on-sky performance is still being evaluated; however, given VISION is largely a visible light analog of the near-infrared MIRC, the anticipation is that the performance should follow MIRC's impressive closure phase precisions.

\section{Potential Planet Transit Targets}\label{sec_potential_targets}

The obvious candidates for observations of a planet transit event are 55 Cnc (\cite[McArthur et al. 2004]{2004ApJ...614L..81M}) and HD189733 (\cite[Bouchy et al. 2005, Bakos et al. 2006]{2006ApJ...650.1160B,2005A&A...444L..15B}). With angular sizes of $711 \pm 4$ and $376 \pm 31$ $\mu$as, respectively (\cite[von Braun et al. 2007, Baines et al. 2007]{2011ApJ...740...49V,baines2007}), they are the planet-transit hosting stars with the largest angular sizes discovered to date. The notional geometry of the HD189733 transit is depicted in the upper row of Figure \ref{fig_HD189733CP}.

The next best candidate known at the time of this draft is GJ 436, with roughly the same anticipated angular size as HD189733. HD149026, HD17156, and HD209458 are also worth considering, although their stellar angular diameters of 170-250 $\mu$as are significantly less favorable for detection when considering the maximum baselines of CHARA-MIRC (3$\times$330m) and currently available baselines ($\sim 80$m) for NPOI-VISION.  (The NPOI facility has infrastructure out to 430m, which is underoing commissioing and should enable routine observations of stars with diameters down to $\sim$80$\mu$as.)

\section{Transit Closure Phase and Visibility Deviations}\label{sec_deviations}

The use of the closure phase observable effectively cancels many of the corrupting effects of
the atmosphere and the instrument, and is a highly sensitive probe for interferometric image construction
on the smallest spatial scales.  A significantly more complete discussion of the topic of
closure phase may be found in \cite{2007NewAR..51..604M}.

Closure phases are very sensitive to asymmetries in
images, which will prove quite useful in the application discussed here.
For a star with a planet blocking out part of its disk during a transit event, the degree
of asymmetry is extreme - such a transiting planet is, in essence, a `perfectly black' star spot.

\begin{figure*}
\includegraphics[scale=0.66,angle=0]{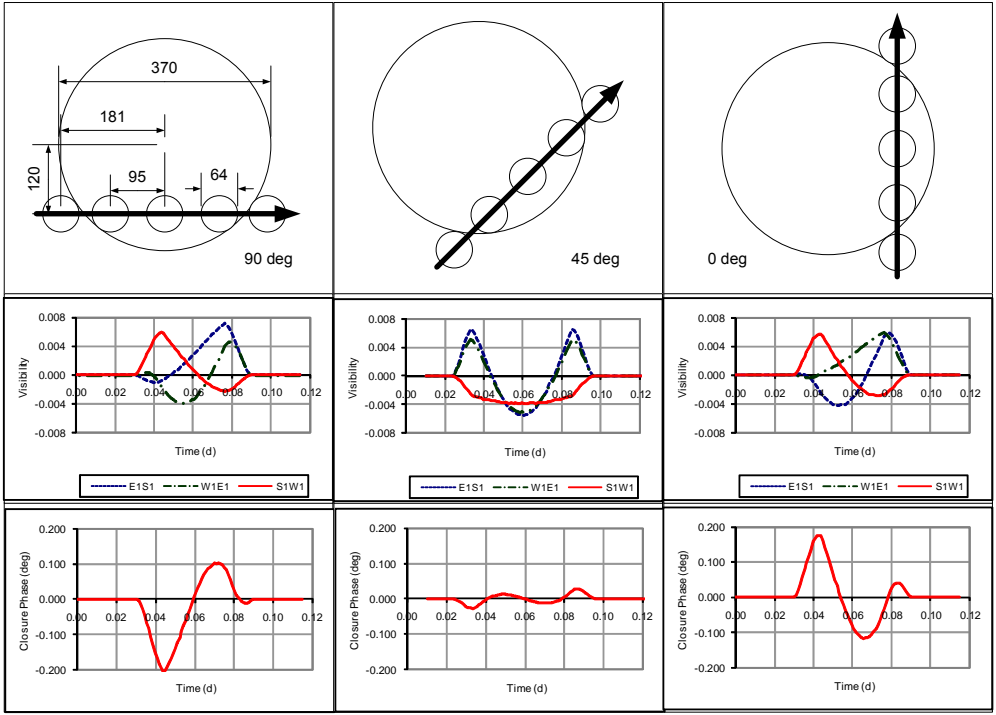}
\caption{\label{fig_HD189733CP} Excursions in visibility amplitude and closure phase data for HD 189733 as observed by CHARA for 3 different possible orientations of the $\sim1^h$ long transit event, as discussed in \S \ref{sec_deviations}. The top row follows Figure 1 and is the image on the sky (units are in microarcseconds), the middle row is the visibility amplitude excursions for each of the 3 longest CHARA baselines (E1-S1-W1), and the bottom row is the closure phase difference for that baseline triangle during the transit event.}
\end{figure*}

The full envelope of expected visibility amplitudes and closure phases
for a {\it gedanken} experiment covering the interferometer response
can be computed for a a planet transit event.
For the specific case of HD189733, such an experiment using a notional CHARA E1-S1-W1 configuration may easily be
executed, and is discussed in detail in \cite[van Belle (2008)]{2008PASP..120..617V}, leading to an expectation of the changes in observed visibility amplitudes and
closure phases during the planet transit event (Figure \ref{fig_HD189733CP}).

During the transit event,
the visibility deviates from from the nominal unocculted star case, but
by only a marginal amount - on the order of $\pm0.01\%$, which is
beyond the capabilities of any existing interferometer by $\sim$two orders of magnitude.  However, the
closure phase excursion is $\pm0.2^o$; as detailed in \cite{2006SPIE.6268E..55M} and \cite{2011PASP..123..964Z}, closure phase
measurements at this level of precision appear possible.  Tests of the CHARA-MIRC
system showed laboratory closure phase formal error at the $\sigma_\Phi \sim 0.03^o$ level over the course of 3 hours.
Shorter integration times indicated a correspondingly higher level of scatter, but the magnitude
of this error gives a starting point from which to evaluate the possibility for observation
of a planet transit event using closure phases.

\section{Discussion and Conclusion}

With a crude level of closure phase error ($\overline{\sigma_\Phi} \sim 0.1^o$), the position angle of the transit event is readily recovered to within a few degrees, and markedly better with modest improvements in closure phase error.  Since closure phase is, in essence, an observable that quantifies the degree of asymmetry in an image upon the sky, we expect that this technique should work well even for grazing transit events.
Of particular interest is the fact that, in the best conceivable case for each apparent transit position angle, the planet radius appears to be recoverable to the level of roughly one part in 30-40, which appears to beat the previous interferometric measure of the planetary diameter (\cite[Baines et al. 2007]{baines2007}) by a factor of 2.5$\times$.
This approach is the best currently available technique that provide any value for the transit event orientation angle; it is also an independent check on parameters such as stellar radius or planet-star radius ratio that is derived from other techniques, such as spectroscopy or photometric timing.

Finally, it is worth noting that an interesting test of this approach would be stellar disk imaging during occultations by (local) asteroids.  Such events are rare for specific objects, but an ensemble search of the $>$400,000 known asteroids' paths intersecting the positions of the $>$10,000 stars bright enough for CHARA-MIRC and NPOI-VISION presents a list of a few dozen events annually.  The key limitation for these events is the duration, which is not measured in hours (eg. exoplanet transits) but in minutes and seconds.

\end{document}